\documentclass[10pt,aps,prl,twocolumn,showpacs,superscriptaddress]{revtex4-1}

\pdfoutput=1

\usepackage[utf8]{inputenc}
\usepackage[linktocpage=true, colorlinks=true, urlcolor=blue, linkcolor=blue, citecolor=blue]{hyperref}
\usepackage{graphicx}
\usepackage{amsmath,amssymb}

\usepackage[english]{babel}

\usepackage{color}
\usepackage{subfigure}
\usepackage{epstopdf} 
\usepackage{url}

\usepackage{xspace}

\begin{document}

\title{One single static measurement predicts wave localization in complex structures}

\author{Gautier Lefebvre}
\affiliation{
\'Ecole Sup\'erieure de Physique et de Chimie Industrielles ParisTech, Paris Sciences et Lettres Research University, CNRS, Institut Langevin, F-75005 Paris, France\\
}
\author{Alexane Gondel}
\affiliation{
Mines Paris-Tech, 60 Bld Saint-Michel, Paris, France\\
}
\author{Marc Dubois}
\author{Michael Atlan}
\affiliation{
\'Ecole Sup\'erieure de Physique et de Chimie Industrielles ParisTech, Paris Sciences et Lettres Research University, CNRS, Institut Langevin, F-75005 Paris, France\\
}
\author{Florian Feppon}
\author{Aim\'e Labb\'e}
\author{Camille Gillot}
\author{Alix Garelli}
\author{Maxence Ernoult}

\affiliation{Physique de la Mati\`ere Condens\'ee, Ecole Polytechnique, CNRS, 91128 Palaiseau, France}

\author{Svitlana Mayboroda}
\affiliation{School of Mathematics, University of Minnesota, Minneapolis, MN, USA}

\author{Marcel Filoche}
\affiliation{Physique de la Mati\`ere Condens\'ee, Ecole Polytechnique, CNRS, 91128 Palaiseau, France}

\author{Patrick Sebbah}
\email[Contact: ]{patrick.sebbah@espci.fr}
\affiliation{
\'Ecole Sup\'erieure de Physique et de Chimie Industrielles ParisTech, Paris Sciences et Lettres Research University, CNRS, Institut Langevin, F-75005 Paris, France\\
}
\affiliation{Department of Physics, The Jack and Pearl Resnick Institute for Advanced Technology, Bar-Ilan University, Ramat-Gan, 5290002 Israel
}

\date{\today}

\begin{abstract}
A recent theoretical breakthrough has brought a new tool, called \emph{localization landscape}, to predict the localization regions of vibration modes in complex or disordered systems. Here, we report on the first experiment which measures the localization landscape and demonstrates its predictive power. Holographic measurement of the static deformation under uniform load of a thin plate with complex geometry provides direct access to the landscape function. When put in vibration, this system shows modes precisely confined within the sub-regions delineated by the landscape function. Also the maxima of this function match the measured eigenfrequencies, while the minima of the valley network gives the frequencies at which modes become extended. This approach fully characterizes the low frequency spectrum of a complex structure from a single static measurement. It paves the way to the control and engineering of eigenmodes in any vibratory system, especially where a structural or microscopic description is not accessible.

\end{abstract}

\pacs{63.50.-x, 63.20.Pw, 43.40.At, 43.40.Dx}

\maketitle


One of the key features exhibited by stationary waves in complex geometry or disordered systems is \emph{localization}, characterized by an unexpected concentration of energy within a small portion of the system even in the absence of any confining potential~\cite{Anderson1958}. Localization of waves may occur in the presence of structural or geometrical heterogeneities. A random potential, distributed scatterers, or specific boundary geometry can induce mode confinement under the right conditions~\cite{Even1999}. Even the simplest geometry can produce localization, as recently observed for mechanical vibrations in rigid plates with a single clamped point~\cite{FilocheMayboroda2009}. 
More complex geometries have demonstrated an efficient way to concentrate and dissipate energy in acoustical cavities~\cite{EPL2009,FractalWall}, to tailor electromagnetic anechoic chambers~\cite{Hemming2002} or to design musical instruments~\cite{Boutillon13}.
In nanophotonics, controlling light confinement is also an important challenge~\cite{Vahala2003}. Indeed geometry of optical cavities, photonic crystals~\cite{Cao2011} as well as disordered structures~\cite{Wiersma2011} can be designed to optimize light trapping~\cite{Noda2005}, miniaturize lasers~\cite{Noda2006}, improve absorption efficiency of thin-film solar cells \cite{Wiersma2012}, store quantized bit of light~\cite{Lannebere2015}, or modify spontaneous emission in cavity quantum electrodynamics~\cite{Sapienza2010,Sapienza2011}. Actually, in most cases, the investigation of localization is either based on empirical knowledge, numerical simulations or optimization algorithms~\cite{Mosk12,Bachelard2014}.

Being able to systematically predict the spatial and spectral characteristics of the confined modes and to address the question of where the modes will localize remain therefore a major challenge, and often requires solving for a given geometry the full eigenvalue problem. In this letter, we present a much simpler and more universal approach, taking advantage of the landscape theory~\cite{FilocheMayboroda2012}. To this end, we design a test structure, a thin plate with complex geometry which exhibits localization of flexural waves. We show that the landscape function is a physical quantity directly accessible to the measure. Using laser holographic heterodyne technique, we measure the static deformation of the plate under uniform load as well as its vibration modes. The ``valley network'' of the landscape precisely corresponds to the localization regions of the modes and the ``hill peaks'' give a good estimate of the corresponding resonance frequencies. Finally, the opening of the localization regions at higher frequencies is successfully predicted by the measured landscape as well and is confirmed by the observation of the transition from confined to extended modes.
Because of the key information it contains, the landscape opens new perspectives in terms of measurements and design of complex vibrating systems.

\begin{figure}[]
\centering
\includegraphics[width=0.5\textwidth]{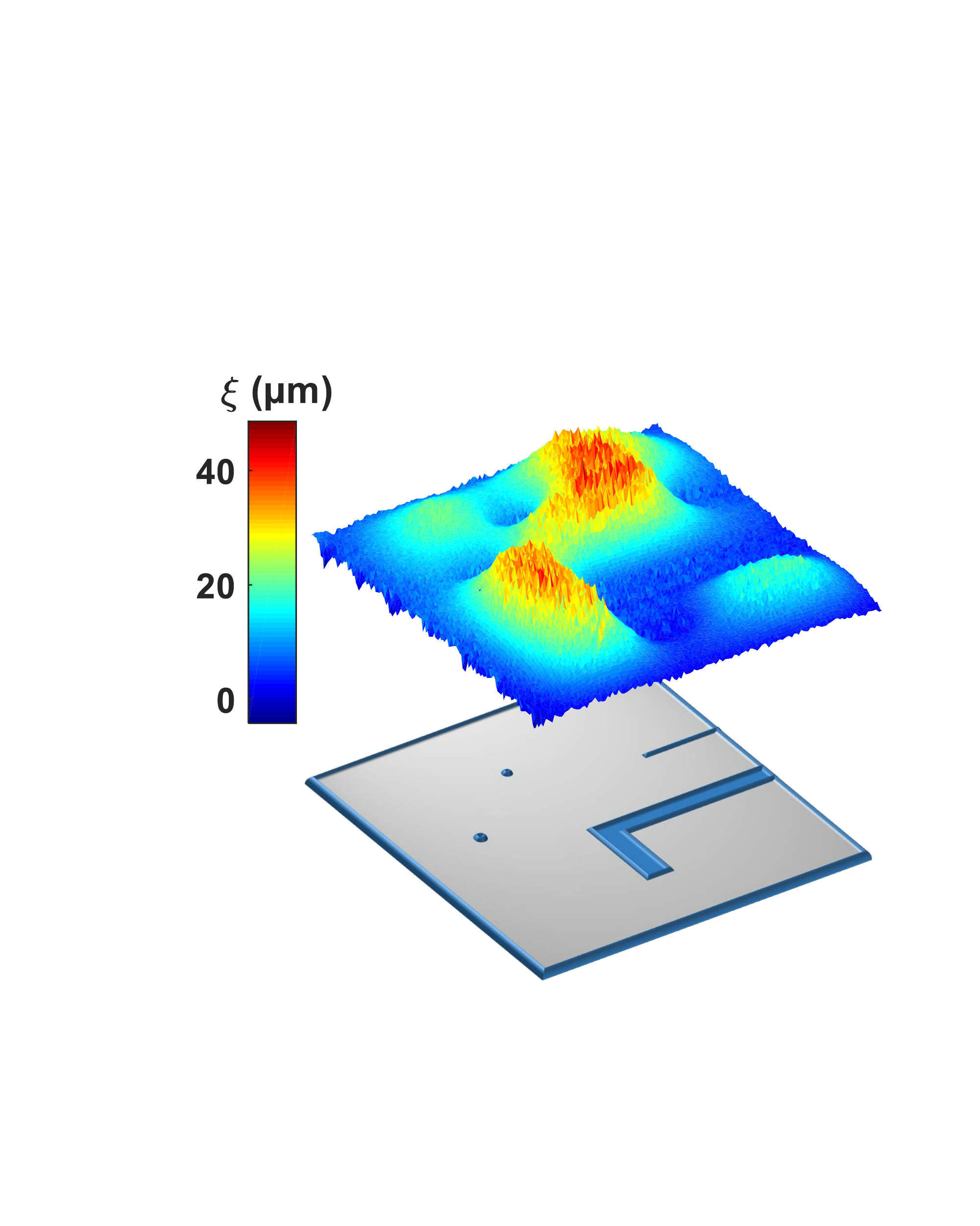}
\caption{(color online).\textbf{The localization landscape.} (top):~Absolute static deformation in $\mu$m of the 2D rigid thin plate under uniform pressure obtained by measuring the holographic phase map accumulated as pressure exerted on the plate is released (see Movie 1 in Supplemental Information \cite{SM}). (bottom):~Drawing of the 10 cm $\times$ 10 cm-large, 0.5 mm-thick metallic plate. The blue regions represent schematically the clamped regions: the edges, a L-shape carved region, a segment and two points. The actual design of the system is detailed in the Supplemental Information \cite{SM}).}
\label{fig_landscape3d}
\end{figure} 

The system that we investigate is shown in Fig.~\ref{fig_landscape3d}. It consists of a 10~cm-side Duraluminium square plate, 0.5 mm thick, with edges and different regions clamped, including a L-shape carved region, a segment and two points. A detailed description of how the system is designed is provided in the Supplemental Information \cite{SM}). This geometry is inspired from the system studied in~\cite{FilocheMayboroda2012}, where it was shown that its modal spatial distribution is highly non intuitive as it depends on the nature of the system (e.g. a rigid plate or a membrane) and, respectively, the associated canonical equation (bi-Laplacian or simple Laplacian). 

We assume that the vibrating plate is rigid and thin compared to the wavelength, and that a local excitation results essentially in the creation of steady-state resonances of 0-order anti-symmetric ($A_0$) Lamb waves, also called flexural waves in the low frequency regime. In that limit, the wave motion in an isotropic thin plate is well approximated by the Kirchhoff–Love equation~\cite{BookRoyerDieulesaint2000}
\begin{equation}\label{eq_PlateFlexuralMotion}
\frac{\partial^2w}{\partial t^2} + \frac{Eh^2}{12\rho\left(1-\nu^2\right)}\Delta^2w = 0 \quad,
\end{equation} 
where $w(x,y)$ is the out-of-plane displacement, $h=0.5$~mm is the plate thickness, and $E=72.5$~GPa, $\rho=2.79$~g$\cdot$cm$^{-3}$, and $\nu=0.33$ are respectively the Young's modulus, the density, and the Poisson ratio of the material (here Duraluminium). The harmonic solutions of Eq.~\ref{eq_PlateFlexuralMotion} take the form 
\begin{equation}\label{eq_SteadyStateSolutions}
w(x,y,t) =~ W\left(x,y\right)~\exp( i \omega t )
\end{equation}
The Rayleigh-Lamb dispersion relation of the $A_0$ mode in the low frequency approximation~\cite{BookRoyerDieulesaint2000} relates the acoustic wave number~$k$ to the frequency~$\omega$:
\begin{equation}\label{eq_DispersionRelationship}
k^2 = \alpha \omega \quad \mbox{with} \quad \alpha = \sqrt\frac{12\rho\left(1-\nu^2\right)}{E h^2}~.
\end{equation}
The eigenmodes of vibration~$W_m$ at the angular frequencies $\omega_m$ therefore satisfy the steady-state equation of flexural waves~\cite{TrefethenBetcke2006} derived from Eq.~\ref{eq_PlateFlexuralMotion}, Eq.~\ref{eq_SteadyStateSolutions}, and Eq.~\ref{eq_DispersionRelationship}. The edges of the plate are clamped to be motionless, which results in vanishing vibration amplitude and spatial derivative at the boundaries. Calling respectively $\Omega$ and $\partial \Omega$ the plate and its boundary, the mathematical formulation of the problem finally writes
\begin{eqnarray}
\label{eq_E}
\begin{cases}
L~W_m = \omega_m^2~W_m \quad &\mathrm{on}~\Omega,\\
W_m = 0 \quad &\mathrm{on}~\partial \Omega,\\
\partial_\nu W_m = 0 \quad &\mathrm{on}~\partial \Omega,
\end{cases}
\end{eqnarray}
where $L$ is the elliptic operator 
\begin{equation}\label{eq_EllipticOperator}
L = \frac{1}{\alpha^2}\Delta^2
\end{equation}
and $\partial_\nu$ is the normal derivative.
According to the new localization theory proposed in~\cite{FilocheMayboroda2012}, most of the information on flexural wave localization can in fact be retrieved from a mathematical object called the \emph{localization landscape}. This object is a positive function~$u(x,y)$ defined as
\begin{equation}
\label{eq_udef}
u(x,y) = \int_\Omega |G(x',y':x,y)|~dx'dy'
\end{equation}
where $G$ stands for the Green function of the associated wave operator~$L$. The landscape function controls the amplitude of the localized waves in the entire domain, which implies that the regions of low values of $u(x,y)$ are also the regions of small vibratory amplitude. In other words, the curves where $u$ is small (refered to as the ``valleys'' hereafter) produce invisible barriers for waves. The propensity of the landscape $u$ to constrain the amplitude of a steady-state vibration~$W_m$ emerges through the following inequality~\cite{FilocheMayboroda2012}
\begin{equation}
\label{eq_LandscapeConstraint}
|W_{m}(x,y)|~\leq~\omega_m^2~u(x,y) \quad \forall (x,y) \in \Omega
\end{equation}
where $W_m$ is normalized so that the maximal amplitude is equal to 1. Due to this specific choice of normalization, Ineq.~\ref{eq_LandscapeConstraint} corresponds to an actual constraint on the mode amplitude only at the points where $\omega_m^2 u(x,y) < 1$. In this picture, the valleys of $u$ 
delimit the confining subregions for the localized eigenmodes. In summary, the partition of the plate created by these lines enables us to predict the subregions of the plate $\Omega$ where the vibrations will be localized.

The definition of~$u$ given in Eq.~\ref{eq_udef} makes it a complicated quantity to compute or to measure in general. However, if one assumes that the Green functions are positive everywhere (a property almost satisfied by the bi-Laplacian operator, and that we will discuss later in the article), then the absolute value can be removed in Eq.~\ref{eq_udef}. In this case, $u$ becomes the solution of the following Dirichlet problem: 
\begin{eqnarray}
\label{eq_DirichletLandscape0}
\begin{cases}
L~u = 1 \quad &\mathrm{on}~\Omega,\\
u = 0 \quad &\mathrm{on}~\partial \Omega,\\
\partial_\nu u = 0 \quad &\mathrm{on}~\partial \Omega.
\end{cases}
\end{eqnarray}
In physical terms, the landscape $u$ is thus the out-of-plane static deformation $\xi$ under uniform load, modulo a multiplicative constant, that is
\begin{equation}
\label{eq_displacement}
u = \frac{\rho h}{P_0}~\xi~,
\end{equation}
where $P_0$ is the applied pressure on the plate. This property has a very important consequence: it means that, without any computation or a priori knowledge of the system,  the direct static measurement of $\xi$ brings geometrical information about the localization subregions and quantitative information about the threshold above which delocalization will occur. This can prove particularly useful in disordered or random systems where the microscopic and structural information is lacking or unattainable.

\begin{figure}[]
\centering
\includegraphics[width=0.52\textwidth]{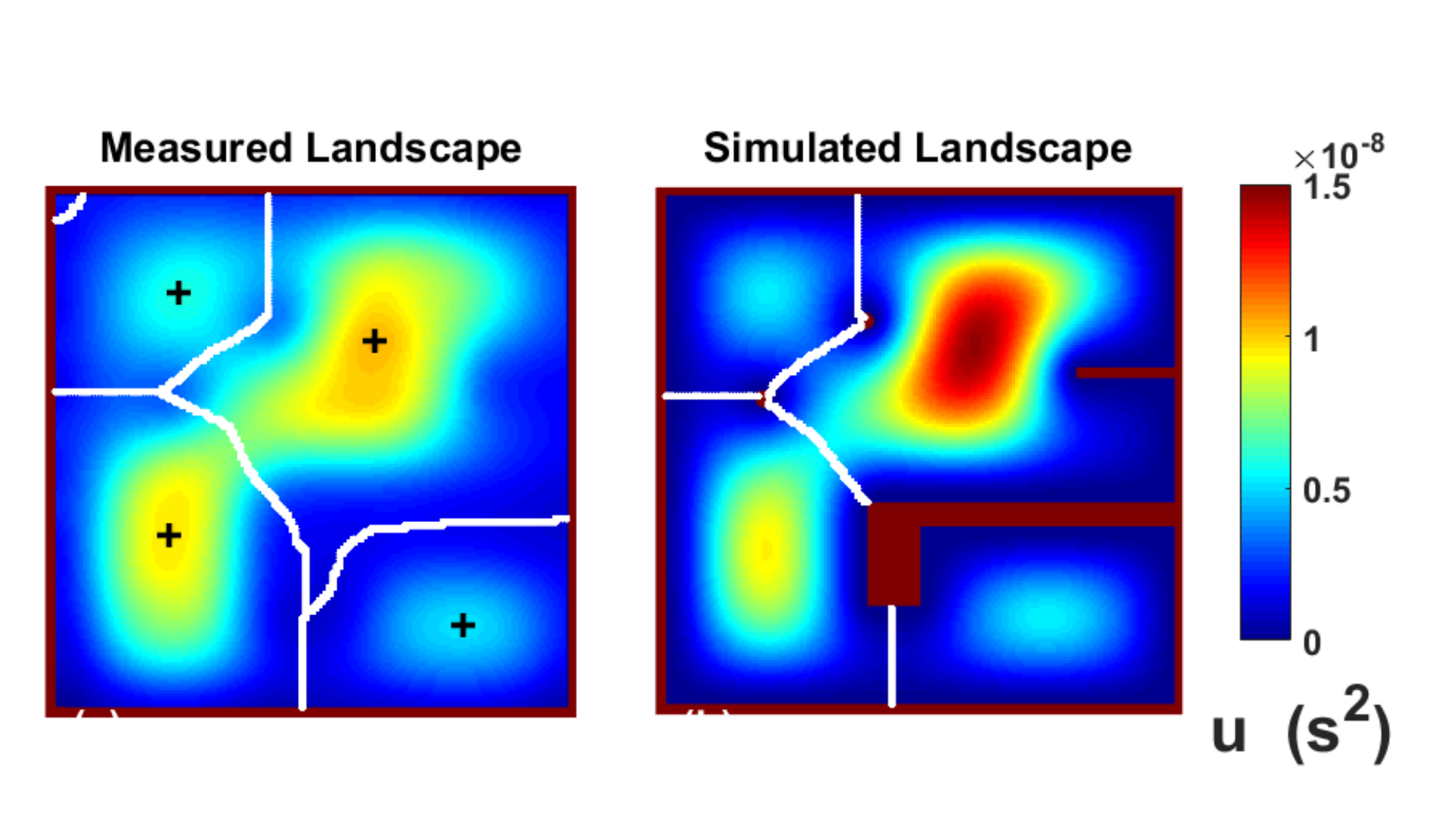}
\caption{(color online). \textbf{Measured vs. simulated landscape}. (a):~2D representation of measured landscape, $u(x,y)$, issued from plate static deformation shown in Fig.~\ref{fig_landscape3d} after data processing (see Supplemental Information \cite{SM})) and applying Eq.~\ref{eq_displacement}. The white lines are the watershed lines of steepest ascent or steepest descent that form the valley network of the landscape. The black crosses indicate the location of the local maxima of $u(x,y)$, from which the frequencies of the lower-spectrum localized modes are determined (see Fig.~\ref{fig_FreqPrediction}). (b):~Calculated landscape using finite element method to solve Eq.~\ref{eq_DirichletLandscape0} for the same rigid plate as in the experiment.}
\label{fig_Landscape}
\end{figure}

To assess this conjecture, we measure the static deformation of the plate under uniform pressure, $P_0$=4600 Pa (as described in the Supplemental Information \cite{SM})) and we obtain the localization landscape shown in Fig.~\ref{fig_landscape3d}a and Fig.~\ref{fig_Landscape}a ((see also Movie 1 in Supplemental Information \cite{SM})). We can see the influence of the clamped regions on the landscape, constraining the amplitude of the deformation to remain smaller in their vicinity while other regions of the plate sustain a larger stretching. From this direct measurement, 4~local maxima are detected, hence 4~localization subregions.

This measurement is compared to a numerical simulation of the landscape function solution of Eq.~\ref{eq_DirichletLandscape0}, computed using a finite element method (FEM)~\cite{HechtPironneau2005} and assuming a homogeneous medium (Fig.~\ref{fig_Landscape}b). The network of valleys is almost identical to the one measured. It partitions the plate into the same 4~domains.

\begin{figure}[]
\centering
\includegraphics[width=0.5\textwidth]{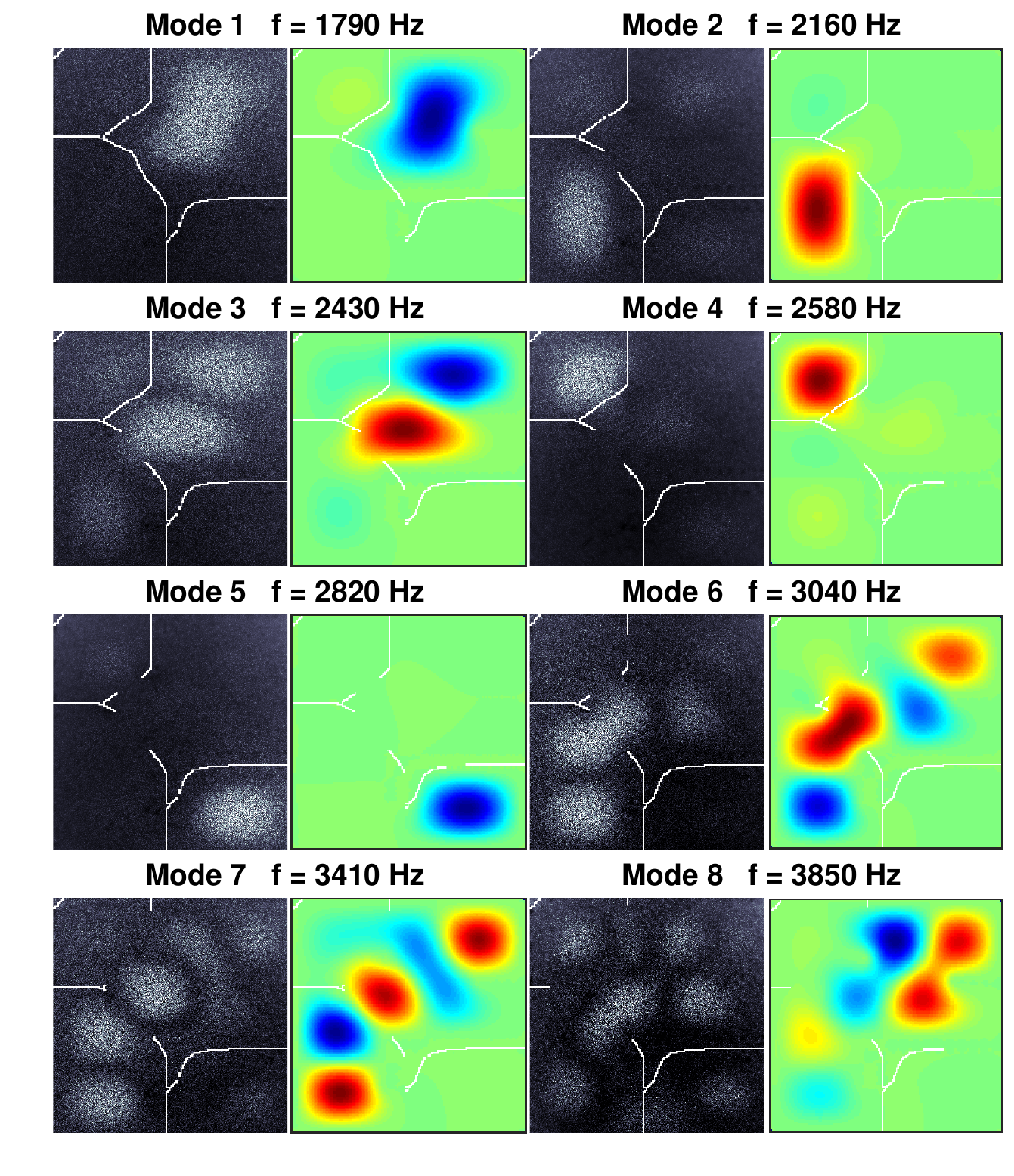}
\caption{(color online). \textbf{Vibration modes}. Local amplitude of flexural waves at the eight first resonances. Experimental optical measurements (grey scale) using heterodyne holographic interferometry are compared side-by-side with finite element method calculation (color scale) of the full elastic problem. Experiments and Numerics give identical resonance frequencies within the measure precision. The superimposed white lines represent a subset of the valley network where $u(\vec{x}) \le \omega^2$ and delimit the localization areas. As frequency increases, gaps open in the valley network and modal confinement constraint is released.}
\label{fig_modes}
\end{figure} 

Using narrow-band and wide-field imaging heterodyne optical holography (see  Supplemental Information \cite{SM}) and scanning over frequency, we measure the spatial distribution of the out of plane vibration at resonance. The first eight modes are displayed in Fig.~\ref{fig_modes} and compared to FEM numerical simulations. The amplitude distribution of the measured eigenmodes appear very similar to the calculated vibration pattern at resonance. 

The valley lines of the landscape are superimposed on each plot of Fig.~\ref{fig_modes}. For a given frequency~$\omega$, only the portion of these lines where $\omega^2 u(x,y) < 1$ is plotted. Due to the normalization of the mode amplitude in eq.~\ref{eq_LandscapeConstraint}, this subset of the valley lines is the only one exerting an actual control on the mode amplitude (see Eq.~\ref{eq_LandscapeConstraint}). We observe that the first eigenmodes are confined within the localization subregions predicted by the static measurement of the localization landscape. In fact, at low frequency, each eigenmode is almost an eigenfunction of one of the localization subregion.

\begin{figure}[]
\centering
\includegraphics[width=0.4\textwidth]{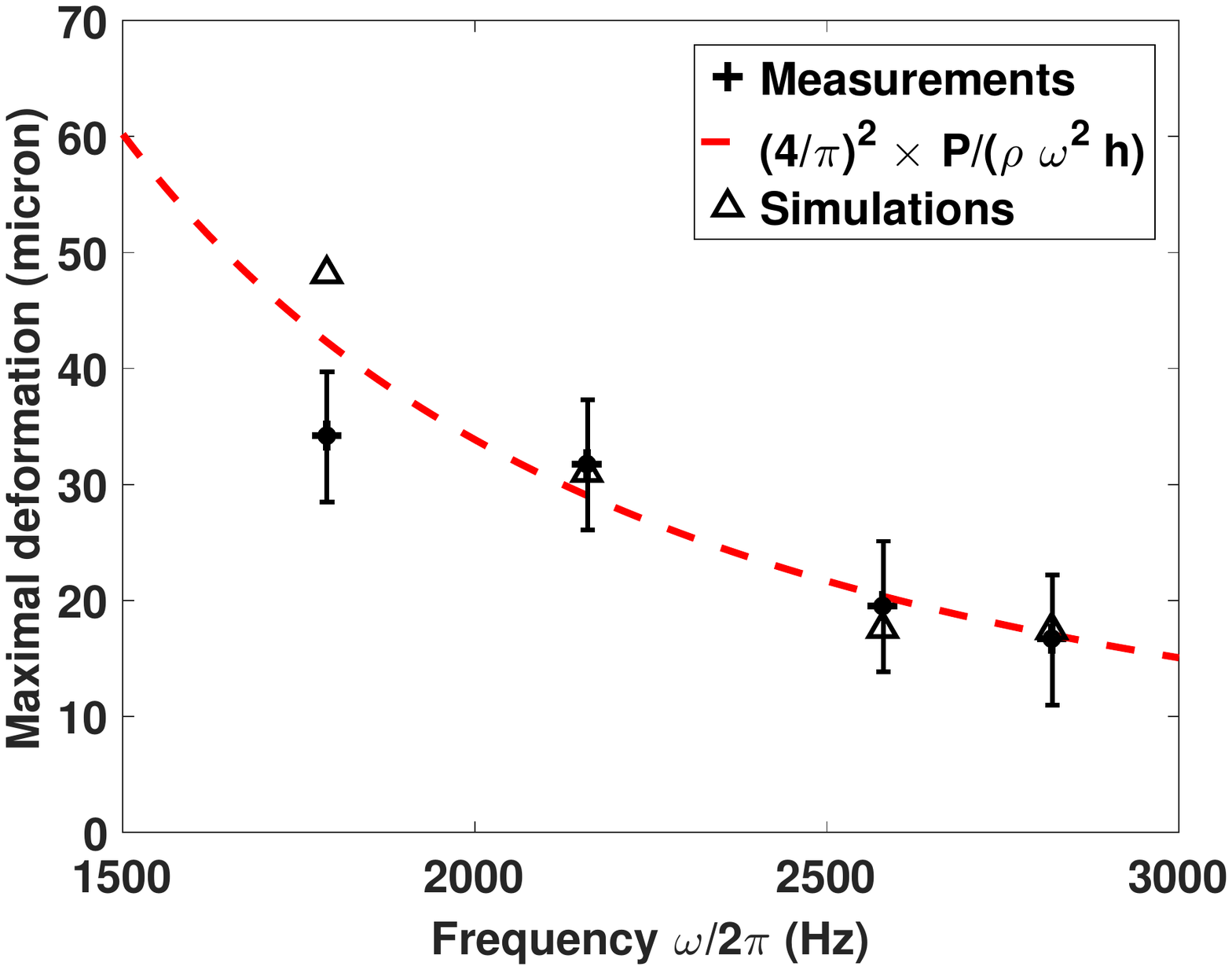}
\caption{(color online). \textbf{Predicting the low frequency spectrum}. \textbf{Crosses}:~Measured local maxima of the plate static deformation in each localization subregion (shown by crosses in Fig.~\ref{fig_Landscape}) vs. measured resonance frequencies of the first modes localized in the corresponding regions (mode 1,2,4 and 5 of Fig.~\ref{fig_modes}. The error bars reproduce the uncertainty in the phase reference (see Supplemental Information \cite{SM})). \textbf{Open triangles}: ~same obtained from numerical simulations. \textbf{Dashed line}:~Theoretical prediction (see Supplemental Information \cite{SM})). }
\label{fig_FreqPrediction}
\end{figure}

Going further, the landscape provides not only spatial but also spectral information. Indeed, the resonance frequency of the first modes can be estimated directly from the landscape function~$u(x,y)$. We show in the \cite{SM} that theoretically $\omega_i \simeq 1.27/\sqrt{\max_i(u)}$, where $\omega_i$ is the frequency of the fundamental mode (one single peak) in subregion $i$, and $\max_i(u)$ is the local maximum of $u$ in the same region. This relation is confirmed experimentally by plotting in Fig.~\ref{fig_FreqPrediction} the measured resonance frequencies of modes 1, 2, 4 and 5 vs the local maxima of the landscape function in the corresponding localization subregions. Remarkably, experimental points fit the theory within the error bars (see Supplemental Information \cite{SM})). Frequency estimate for higher order modes with multiple bumps is also possible, but requires a more involved analysis. This is currently work in progress. The results presented here are the first step towards a prediction of the entire spectrum of vibratory systems from a simple static measurement.

As the frequency increases, the constraint expressed in Eq.~\ref{eq_LandscapeConstraint} loosens, and gaps open along the valley lines ((see Movie 2 in Supplemental Information \cite{SM})). As a result, initially isolated subregions become connected, meaning that modes can extend over larger domains. This is illustrated in Fig.~\ref{fig_modes} where e.g. mode~2 remains confined while mode~3 brims over the small gap newly opened. Total mode delocalization over two subregions that were initially disconnected is observed e.g. for mode~6. As the gaps widen, larger localization regions are formed and modes extend over the entire system. The frequency~$\omega$ at which a gap opens along a valleyline $\Gamma$, satisfies $\omega^2 \max_\Gamma(u) = 1$, where the maximum of~$u$ is taken over $\Gamma$. This maximum is directly retrieved from the measured landscape function $u(x,y)$. 

We have seen earlier that the positivity of the Green functions of the wave operator leads to system~\ref{eq_DirichletLandscape0} satisfied by the landscape function. One has to point out that, in all generality, the bi-Laplacian with Dirichlet boundary conditions (vanishing amplitude and vanishing normal derivative) is not a positive operator. It means that the solution to a Dirichlet problem with positive load may change sign~\cite{GrunauSweers2014}. For example, it has been shown mathematically that the deformation of a square plate under uniform load exhibits an infinite number of smaller and smaller oscillations near the corners of the plate~\cite{Coffman1982}. However, the amplitude of these alternating oscillations is so tiny that it is not measurable in a practical experiment. This is the case for most mechanical thin plates~\cite{Gazzola2010}. Thus, in practice, the solution to the problem~(\ref{eq_DirichletLandscape0}), obtainable from one static measurement only, is extremely close to the theoretical function defined in~Eq.~\ref{eq_udef}.

These results demonstrate experimentally the predictive nature of the landscape in a physical situation. The general behavior presented here shows that fundamental vibratory properties of the system are encoded in the landscape function, obtained from one static measurement. Remark that not only the landscape predicts the shape and location of the confining regions, as well as the frequencies of the localized low-energy modes, but it also gives access to the value of the transition frequencies where the progressive coupling between neighboring regions eventually leads to hybridization of their steady-state vibrations~\cite{Labonte2012}. This is a first step in understanding the transition from localized to extended modes regime.

These results can be generalized to any disordered or structured system, where the structural or microscopic information is not accessible, and therefore where no numerical solution of the localization landscape can be computed. They establish a strong and rigorous relationship between the static and the dynamic properties of vibrating systems, independently of their dimensions or the nature of the vibrations. They further demonstrate how the landscape function can grant the experimentalists predictive power on the dynamical behavior of a system without having to force it or to solve the full modal problem. In the next step, the localization landscape should become a tool of choice for addressing the inverse problem, i.e. building the structure with desired spatial and frequency vibratory properties~\cite{Benisty2005}.

\subsection*{Acknowledgments}

The authors thank Dominique Clément for the plate design and realization. P.~Sebbah is thankful to the Agence Nationale de la Recherche support under grant ANR PLATON (No. 12-BS09-003-01), the LABEX WIFI (Laboratory of Excellence within the French Program Investments for the Future) under reference ANR-10-IDEX-0001-02 PSL* and the PICS-ALAMO. This research was supported in part by The Israel Science Foundation (Grant No. 1781/15 and 2074/15). S.~Mayboroda is partially supported by the Alfred P. Sloan Fellowship, the NSF CAREER Award DMS-1056004, the NSF MRSEC Seed Grant, and the NSF INSPIRE Grant. M.~Filoche is partially supported by a PEPS-PTI Grant from CNRS.

\end{document}